\documentclass[aps,twocolumn,superscriptaddress]{revtex4}
\usepackage{graphicx}
\usepackage{dcolumn}
\usepackage{bm}
\usepackage{amssymb,amsfonts,amsmath}
\def\beq{\begin{equation}}
\def\eeq{\end{equation}}
\begin{document}

\title{Diversity Begets Stability in an Evolving Network}

\author{Ravi Mehrotra}
\affiliation{National Physical Laboratory, K. S. Krishnan Rd., New
Delhi-110012, India}
\author{Vikram Soni}
\affiliation{National Physical Laboratory, K. S. Krishnan Rd., New
Delhi-110012, India}
\author{Sanjay Jain}
\email[Corresponding author:]{jain@physics.du.ac.in}
\affiliation{Department of Physics and Astrophysics, University of
Delhi, Delhi 110007, India} \affiliation{Jawaharlal Nehru Centre
for Advanced Scientific Research, Bangalore 560064}
\affiliation{Santa Fe Institute, 1399 Hyde Park Road, Santa Fe, NM
87501, USA}

\begin{abstract}
Complex evolving systems such as the biosphere, ecosystems and
societies exhibit sudden collapses, for reasons that are only
partially  understood. Here we study this phenomenon using a
mathematical model of a system that evolves under Darwinian
selection and exhibits the  spontaneous growth, stasis and collapse
of its structure. We find that the typical lifetime  of the  system
increases sharply with the diversity of its components or species.
We also find that the prime reason for crashes is a naturally
occurring internal fragility of  the system. This fragility is
captured in the network  organizational character  and  is  related
to a reduced multiplicity of pathways between its components. This
work suggests new parameters for understanding the robustness of
evolving molecular networks, ecosystems, societies, and markets.
\end{abstract}

\maketitle

Crashes in complex systems include mass extinctions in the biosphere
as evidenced in  the paleontological   record \cite{Raup}\,
collapses   of    ecosystems \cite{Paine}, civilizations
\cite{Diamond}, and crashes  of  stock-markets
\cite{Schumpeter,Sornette}. The death of a human being due to  old
age is also an example of collapse  of  a  once-robust  complex
system turned fragile. While some of these catastrophic events are
caused  by large external perturbations like  meteorite  impacts,
famines,  wars and infections, for the vast majority  of  them  no
single  dramatic cause can be traced
\cite{Raup,Diamond,Schumpeter,Sornette}.  Here we explore an
alternative hypothesis that the prime  reason  for crashes is a
fragility in the internal organization  of  these  systems that
naturally develops in the course of their evolution,  making  them
vulnerable to small perturbations. Unfortunately, empirical data
characterizing the `internal fragility' or `robustness' of such
systems is scarce. One of the chief problems in collecting data is
not knowing  what to look for; we do not know what system parameters
can  characterize  its poisedness for  a  crash. Hence,  a  key
step  in  identifying  possible signatures of fragility  is  to
construct  theoretical  and  mathematical models of systems that
exhibit repeated  catastrophes  in  the course  of their time
evolution, whose analysis can reveal structural  and  dynamical
features that make them vulnerable to such events. An important
aspect  of a complex system's organizational structure is the
underlying  interaction network of its
components\cite{Watts-Strogatz,    Barabasi-Albert,
Dorogovtsev-Mendes, Bhalla-Iyengar}; hence we need in particular to
study examples of evolving networks that exhibit crashes and
recoveries.

Our model system \cite{JK1998,JK2001} exhibits these phenomena for
an evolving network of interacting populations, with Darwinian
selection and dynamic feedback loops playing an important role in
system evolution. The system consists of $s$ nodes, whose network of
interactions  is specified completely by its adjacency matrix $C
\equiv  (c_{ij}),  i,j  = 1,\ldots,s$. A node may represent a
molecular species in a prebiotic pond. The   model   is motivated by
the    origin    of    life    problem \cite{Dyson,Kauffman-book,
Bagley-Farmer-Fontana},  but   may be more generally valid. The
element $c_{ij} = 1$ if species $j$ `catalyzes'  the growth of
species $i$, and zero otherwise. Also, $c_{ii} = 0$ for all $i$,
corresponding to the  exclusion  of self-catalyzing  species.
Relaxing the above restrictions by allowing links with different
weights and negative signs does not change the qualitative behaviour
of the model.

Using the adjacency matrix, we  write  an  equation  for  the
population dynamics of the species given by
\begin{equation}
\label{eq:eqnpop}
    \dot{y}_i  = \sum_{j=1}^s c_{ij} y_j - \phi y_i.
\end{equation}
Here, $\dot{y}_i$ is the rate of change of the population of species
$i$. The first term on the right takes into account the positive
effect of  all the  species  that  catalyze  species $i$, each  one
having  an  effect proportional to its population. The second term
is  a  constant  mortality term.

Cast in terms of the relative populations,  $x_i = y_i /
\sum_{i=1}^s y_i$, Eq.~{\ref{eq:eqnpop}} implies that
\begin{equation}
\label{eq:dynamics}
    \dot{x}_i = \sum_{j=1}^s c_{ij} x_j  -  x_i \sum_{k,j=1}^s c_{kj} x_j.
\end{equation}
The dynamics described by Eq.~{\ref{eq:dynamics}} flows to a fixed
point in which all $x_i$ become time independent constants.
Technically,  this steady state is just an eigenvector of the matrix
$C$ corresponding to its largest eigenvalue. For a generic
non-negative matrix $C$, it is a unique, global attractor
(independent  of  initial  conditions),  stable  against
perturbations of the $x_i$.

Initially, the matrix $C$ is sparse and drawn  from  the  random
binomial ensemble with on average $m$ links per node (with $m < 1$).
To introduce  evolution  into the model, we note that the pond can
be washed by nearby tides, floods  or storms that can flush out some
of the contents of the  pond.  We  use  the Darwinian dictum
{\em{`Survival of the  fittest'}}  and  impose that  the species
with the lowest relative  population  in  the steady  state  gets
removed from the system \cite{Bak-Sneppen}; we eliminate the
corresponding node along with all its links from the graph. (If
there are more than  one such species, we choose one at random.)
Furthermore,  such  a  fluctuation can bring in new species into the
pond; we assume for  simplicity  that  a single new node gets added
to the graph whose links with the existing ones are made randomly
with the same average connectivity $m$. After each  such
fluctuation, the populations evolve according  to
Eq.~{\ref{eq:dynamics}} with a fixed $C$ to reach a new fixed point,
whereafter the  above  update sequence is repeated.

\begin{figure}
\includegraphics[scale=0.4]{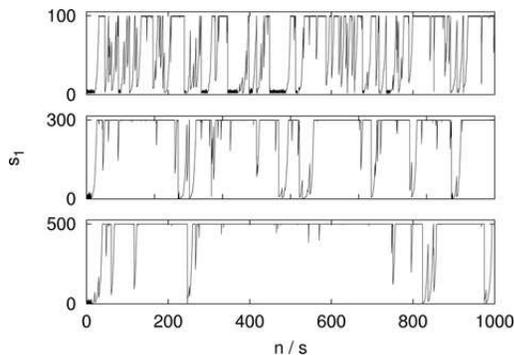}

    \caption
    { Number of populated species $s_1$ as a function of time.  The  total
      number of species $s = 100, 300, 500$ for the three traces from  top
      to bottom, while $m=0.25$ for each. The number of  crashes  decrease
      markedly with increasing $s$.
    }
\end{figure}

\begin{figure}
\includegraphics[trim=200 0 0 0,angle=-90,scale=0.4]{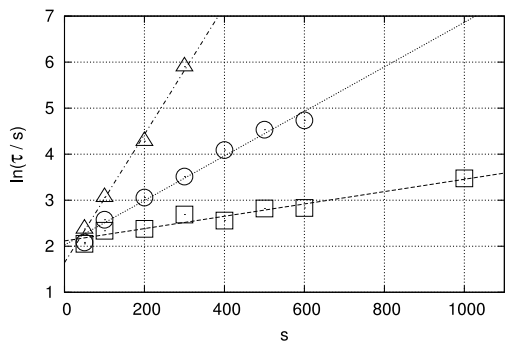}
    \caption
    { Organized state lifetime as a function of $s$ on a semi-log plot for
      $h = 0.75$ and various values of $m$ denoted by $\square :  m=0.15$,
      {\tiny{$\bigcirc$}}~$:m=0.25$, $\triangle :  m=0.35$.  The  straight
      lines are least square fits whose slopes gives $\alpha(m)$.
    }
\end{figure}

At each (`Darwinian')  update,  then,  the  system  suffers  a
structural perturbation that modifies $C$. The perturbation is small
in that only one species is updated, affecting the links of only
$\sim m \sim O(1)$  number of species. Since the update of $C$
depends on populations, the long  time dynamics of the populations
is highly nonlinear inspite of the  simplicity of
Eqs.~{\ref{eq:eqnpop}} and {\ref{eq:dynamics}}. The typical dynamics
is shown in Fig. 1 where the number, $s_1$, of populated species
(whose steady state $x_i > 0$) is plotted against time (number  of
graph updates, $n$) for three values of $s =  100,  300,  500$  and
fixed $m=0.25$. We scale time by $1/s$ as  each  species,  on
average,  can  be updated in $s$ steps. Initially, the graph is
sparse and the number  $s_1$ is small. After a certain time, $s_1$
begins to grow and soon reaches its maximum value $s$. Thereafter,
the system exhibits a stasis for a  certain time scale, $\tau$, in
which $s_1$ fluctuates between $s$  and  $s-1$.  In this state,
which we  call  the `organized  state',  all  species  except
possibly the one being picked for replacement have $x_i > 0$.
Thereafter, the system experiences a collapse in which $s_1$ drops
to  a  fraction  of $s$. This is followed by a recovery and a
repetition of the same kind  of dynamics. This behaviour is
discussed in detail in \cite{JK2002,JK2002a}.

In this letter, we focus on crashes whereby, in a single update
step,  the number of populated species $s_1$ goes from $s$ to a
fraction $h$ of  $s$. We present results for $h=0.50$ and $0.75$.
While the absolute  number  of crashes depends upon $h$, the
qualitative results are not  very  sensitive to its value. As shown
in Fig. 1, for fixed $m$,  the frequency of crashes comes markedly
down with increasing  $s$. Similarly, if we increase $m$ for fixed
$s$, the number of crashes again  decreases markedly.

For given values of $m$ and $s$, there is a typical  lifetime before
the network collapses. We define this time $\tau$  as  the number
of  update steps spent in the organized state in a given run
(typically $10^6$  steps long), divided by the number of crashes
observed during that run. Each run is parameterized by $s$ and $m$
and crashes are defined  with  respect  to the parameter $h$. Hence
$\tau$  depends  upon  $s,  m$,  and  $h$.  The dependence of $\tau$
on $s$ and $m$ is shown in Fig. 2, for $h=0.75$. For fixed $m$,
$\tau/s$ grows exponentially with  $s$.  This behavior is consistent
with the empirical relation \beq
    \frac{\tau}{s} = A(m,h) e^{\alpha(m,h) s}.
\eeq The coefficient $\alpha(m,h)$ is an increasing function of $m$
(and a weak function of $h$) whose quantitative behavior is
discussed later.

These results show that the system is more stable against crashes as
its diversity, $s$, increases for fixed connectivity, $m$, and also
as  its connectivity, $m$, increases for a fixed diversity. We
emphasize  that even for low connectivity the system can be
stabilized  against  collapse by increasing its diversity. It turns
out that in the organized state, the average connectivity of the
species is close to $\sim 1 + m$;  hence,  for the values of $m$
given above the average connectivity  is  only slightly above one.
Even such sparsely connected systems  are stabilized  in this model
by a sufficient amount of diversity.

\begin{figure}
\includegraphics[scale=0.32]{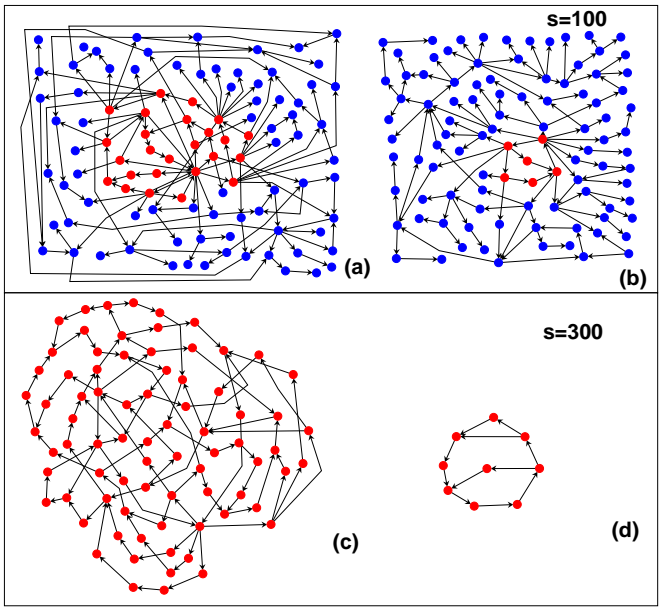}
    \caption
    {Examples of network  configurations  for  $m=0.25$. (a,c): in the
     normal organized state (far away from a crash) and (b,d): in the
     organized state just before a crash. Core nodes are  shown  in  red,
     periphery in blue. For $s$=300, periphery nodes  are  not  shown  to
     avoid clutter. Note (i) that the core  is  large  and  has  multiple
     directed pathways between any  pair  of  its  nodes  in  the  normal
     organized state (a,c). Just before a crash it  becomes  smaller  and
     does not  possess  multiplicity  of  pathways  (b,d).  (ii)  As  one
     increases $s$ from 100 to 300, the number of  multiple  pathways  in
     the core  in  the  normal  organized  state  increases  dramatically
     (compare a and c), while in the state poised for a crash it is  more
     or less the same and quite small (compare b and d).
    }
\end{figure}

We now attempt to understand this behaviour in terms of the
structure  of the graph near and far from a crash. The organized
state has the structure of an {\em{autocatalytic set}} (ACS). An ACS
is a subgraph, each of  whose nodes has at least one incoming link
from a node  belonging  to  the  same subgraph \cite{Kauffman-book}.
In the organized  state,  all  the  species except possibly the one
being picked for replacement are part of the  ACS {\cite{JK1998}}.
The ACS consists  of  a  {\em{core}} and a  {\em{periphery}}. The
core comprises the set of nodes (along with  their  mutual links)
from  which there is a directed path to every other node in the ACS.
All  other  nodes and links in the ACS  constitute the
{\em{periphery}}. Examples of the graph (ACS with core and
periphery) observed in the organized state are shown in Fig. 3. By
definition there is no directed path from a periphery  node  to any
core node. The core, by virtue of closed   paths inside it, is a
`self--sustaining' structure in the sense that all the core nodes
would be populated even if the only links present in the graph are
those  in  the core. In contrast, the periphery nodes would become
depopulated if the links from the core to the periphery were to be
removed. In this sense, the periphery nodes are `parasites' that are
sustained by the core.

While there is always by definition at least one path from every
core node to every other core node,  the  number  of  such  paths is
significantly different between a normal organized state and a state
poised for a crash. In the typical organized state there are several
paths from each core node to another (see Figs. 3(a,c). In  such
configurations, no single node addition or deletion can cause a
crash. However, the number of paths between core nodes drops to a
much  lower  value  just before  a crash (see Figs. 3(b,d)). Then, a
single node change  can disrupt the core and cause most network
species to be depopulated.

In  Fig. 4,  we  plot  the  frequency distribution  of distinct,
non-intersecting closed paths of all lengths in the core in  the
organized state (filled circles). The distribution  shows  a  peak
whose position, $N_p$, is dependent upon  $s$  and  $m$.  A  plot
of  $ln(N_p)$ against $s$ for various values of $m$ is shown in
Fig. 5. This is consistent with the empirical formula \beq
    N_p = B(m) e^{\beta(m) s}.
\eeq We  note  that  loops  in   other   graph   ensembles   have
also   been counted \cite{Bianconi-Marsili}.

In Fig. 4, the open symbols show  the distribution  of closed paths
in the core just before crashes. Its peak occurs  at  a  much
smaller value than $N_p$ (note that the $x$-axis  scale  is
logarithmic). This is also evident from Fig. 3 (the cores  in  (b)
and (d) have much fewer closed paths than in (a) and (c)).

\begin{figure}
\includegraphics[angle=-90,scale=0.4,trim=200 0 0 0]{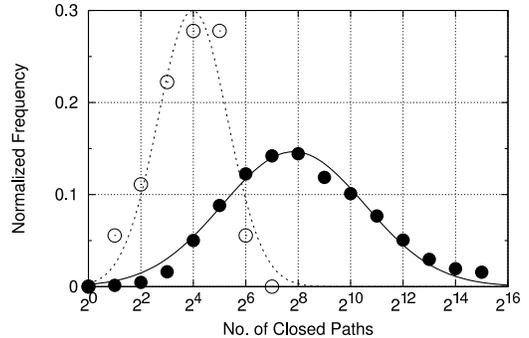}

    \caption
    { Normalized frquency distribution of closed paths in the core  across
      the  sample  of  all  time  steps  in  the  organized  state,   with
      logarithmic   (base   2)   binning,   for   $m=0.25$   and   $s=500$
      ({\LARGE{$\bullet$} } and  solid  line).  Similar  distribution  but
      across the smaller sample  of  time  steps  just  preceding  crashes
      ({\scriptsize{$\bigcirc$}} and dashed line). The  curves  are  least
      square fits to the data using a normal distribution.
    }
\end{figure}

\begin{figure}
\includegraphics[angle=-90,scale=0.3]{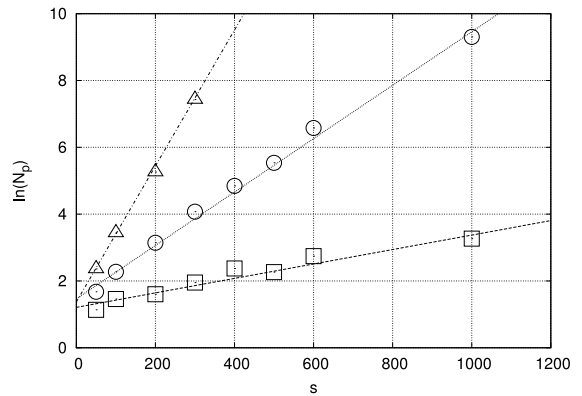}
\caption
    { $ ln( N_p )$ vs. $s$ for various values of $m$  denoted  by  symbols
      $\square : m=0.15$, {\scriptsize{$\bigcirc$}}~$:m=0.25$,  $\triangle
      : m=0.35$. The straight lines are least  square  fits  whose  slopes
      give $\beta(m)$.
    }
\end{figure}

\begin{figure}
\includegraphics[angle=-90,scale=0.4,trim=200 0 0 0]{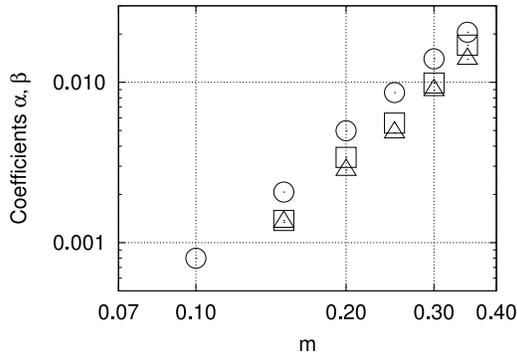}
    \caption
    { A double logarithmic plot of coefficients $\alpha(m)$ and $\beta(m)$
      as functions of $m$. Symbols used are $\square : \alpha(m)$ for $h =
      0.50$,   $\triangle   :   \alpha(m)$   for   $h   =    0.75$,    and
      {\scriptsize{$\bigcirc$}}~:~$\beta(m)$.
    }
\end{figure}

We find a strong correlation  between  the  coefficients $\alpha(m)$
and $\beta(m)$. This is shown in Fig. 6 where $\alpha(m,h)$ for two
values of $h$ and $\beta(m)$ are plotted against $m$. It  is  seen
that the dependence of $\alpha$ on $h$ is weak, as mentioned
before,  and that $\alpha$ and $\beta$ have a similar dependence
upon $m$. Thus  $N_p$ and $\tau/s$ have  a  similar  dependence  on
$m$ and  $s$.  This  close correspondence between a structural
property like the number of  loops  in the graph in the  organized
phase  and  a  dynamical  property  like  the lifetime of that
phase, is one of the surprising results we have found.

This suggests an explanation of why a  higher  diversity  and
density  of links  enhances  stability  against  crashes  in  this
model.   Diversity increases the number of closed paths in  the core
and  thus  provides  a buffer against crashes by ensuring alternate
routes of sustenance  in  the event of loss of core nodes. Crashes
occur typically  when  the  core  has thinned out, and such fragile
states take longer to be realized when there is a larger number of
paths in the core to begin with.

As in real evolutionary systems, the  model  generates  several
dynamical time scales. The model has only two parameters: system
size or  diversity, $s$, and the average connectivity of a new node,
$m$  (the  latter  being typically $O(1)$). In spite of its extreme
simplicity,  the  time  scales that dynamically appear have a wide
range of dependence on $s$,  including logarithmic, power law and
exponential. The time scale for the  appearance of an ACS is
independent of  $s$  and  of  its  growth  is  $\sim \ln  s$
\cite{JK1998,JK2001} (at constant $m$, in scaled units of time as
used  in Fig. 1. Once a crash sets in it occurs fast -- on  a time
scale $\sim 1/s$ in the  present version  of  the  model.  The  fast
collapse and a relatively slower growth seen in the model is  an
observed feature in fossil records as well as stock markets. The
lifetime  of  the system between its growth and collapse has turned
out to be the timescale that is the most sensitive to its diversity,
namely, $\sim e^{\alpha  s}$, as shown here. Such a dependence means
that there is a threshold scale  of diversity set by $1/\alpha$,
such that if diversity  is  well  above  this scale the system is
robust to crashes, but if it is close to or  lower  it is
vulnerable.

The dynamics of growth and collapse in our model is different from
other existing models, including various models of extinction
studied  in  the literature (see the review \cite{Newman-Palmer} and
references  therein). The seed for the growth of complexity in this
model is  a  small  feedback loop (usually a 2-cycle) that  arises
in  the  network  by  chance. The cooperativity implicit in this
autocatalytic structure causes its nodes to have much higher
populations than other nodes. Under a selection dynamics that
preferentially  preserves  nodes  with higher population,  such  a
structure is stable and grows in complexity until  it  spans  the
whole system. Then, the same selection dynamics causes its
components, erstwhile cooperators, to become competitors. This leads
to internal  organizational restructuring, and, on a certain
timescale,  when  the  internal feedback loops become sparse, to
fragility. Thus, we have here an example  of how the very success
and domination  of  a  certain organizational structure changes the
effective rules of the game leading to  the collapse  of  the
structure    (for    another such     example see     the     model
\cite{Cohen-Riolo-Axelrod}).  This   is reminiscent   of   how
certain civilizations   and   organizations collapse   in   the real
world \cite{Diamond}. The role of feedback loops in a network
structure that  evolves  under  both selection  and stochastic
forces   is   also characteristic of several real evolutionary
systems.

Mathematical models of ecosystems suggest that a large number  of
complex factors determine ecosystem stability under various types of
perturbations (see  the   reviews   \cite{McCann, Montoya-Pimm-Sole,
McKane-Drossel, Pascual-Dunne-Levin} and references  therein).  The
importance  of  the multiplicity of sustenance pathways of  species,
suggested  by  MacArthur \cite{MacArthur}, is analogous to the
result we  have  found above.  Note that as in the core of our
graphs, so in  ecosystems at  the  most  basic level there exist
several feedback  loops between  plants  and  microbial communities
that feed  on detritus and  restore  soil  nutrients.  These
self-sustaining parts of the ecosystem are probably  more
primitive,  and their dynamics relatively independent  of  the  more
`peripheral' higher trophic levels that they support. Disruption of
these feedback  pathways would, beyond a certain point, be
catastrophic for  the  ecosystem as  a whole. Most ecosystem models
concerned with stability typically take  into account only the
plants and higher trophic levels  and  exclude  microbes that
provide essential feedback loops. Our work suggests  that  newer
and perhaps clearer patterns  may  emerge  when  models  and  field
data  are considered that include microbes along with other trophic
levels.

A point of caution is that  Eqs.  (1)  and  (2),  motivated  by
catalytic chemical production, would need to be modified to
represent other systems, e.g.,  the  population  dynamics  in
ecosystems.  Nevertheless,   it   is worthwhile to note that several
qualitative  features  of  the  above  model, including
self-organization and collapse of the network, are preserved when we
include negative links (that inhibit species production) and links
with varying strengths, and where the network size $s$ is itself a
dynamical variable with its upper limit statistically determined by
a relative  population threshold  below  which species   are
eliminated \cite{JK2002,KJunpublished,Sandeep-thesis}. The
determination of system lifetime before a crash and the core
architecture in these computationally more demanding versions of the
model is presently under study and will be reported elsewhere.
Needless to say, the real causes of fragility would  be many and
varied for different systems: organisms, ecosystems, societies, etc.
It would be interesting to  explore more realistic models exhibiting
crashes and recoveries to see the  extent to which they share the
behaviour of our simple idealized model.

As in several real world systems, an impending collapse is not
visible  in an obvious way beforehand in the model. The fragility of
the  system  is directly visible only if one looks at the
organizational structure, or the network, and observes the internal
multiplicity  of  core  pathways  (see, e.g., Fig. 3). Thus, it may
not  be  enough  to quantify populations of species in ecosystems,
or  stock  prices of  companies  in markets, or the performance of
individual organs in an aging  human  body. One may need more
`systemic' information about the network of  interaction among the
components and the analysis of internal pathways to  assess  the
true health of these systems.

A notion of robustness of a complex system to the  removal  of nodes
has been defined in ref. \cite{Albert-Jeong-Barabasi} in terms of
the increase of the network diameter. Our approach is different in
that we do not {\em define} system robustness in terms of a network
property, but rather  {\em find} that its robustness as measured by
the time interval between crashes is correlated with a network
property -- the number of closed paths in its core.

It is important to note that our model is  concerned  primarily not
with stability under perturbations of  populations  in  a random
network,  as discussed by May \cite{May1972,May1973}, but with
structural perturbations of node/link removal and introduction that
arise in the natural course  of evolution in a highly self organized
network. It shows how an increase  in diversity and link density can
contribute to long  term  system stability against crashes caused by
such perturbations by increasing the cooperative routes of
sustenance in the network.

\begin{acknowledgments}
We thank Sandeep Krishna for collaboration during the initial phase
of this work and  Areejit  Samal  for  help  with  graph
visualization. S.J. acknowledges support from the Robustness
programme  of the Santa Fe Institute.
\end{acknowledgments}

\end{document}